# Van der Waals trilayers and superlattices: Modification of electronic structures of MoS$_2$ by intercalation

Ning Lu[1,2,⊥], Hongyan Guo[3,2,⊥], Lu Wang[3,4], Xiaojun Wu[3,5], Xiao Cheng Zeng[2,5,*]

[1]Department of Physics, Anhui Normal University, Wuhu, Anhui, 241000, China, [2]Department of Chemistry and Department Mechanics and Materials Engineering, University of Nebraska-Lincoln, Lincoln, NE 68588, USA, [3]CAS Key Lab of Materials for Energy Conversion, Department of Materials Science and Engineering, University of Science and Technology of China, Hefei, Anhui 230026, China, [4]Department of Physics, University of Nebraska-Omaha, Omaha, NE 68182, USA, [5]Hefei National Laboratory for Physical Sciences at the Microscale, University of Science and Technology of China, Hefei, Anhui 230026, China

**Abstract**

We perform a comprehensive first-principles study of the electronic properties of van der Waals (vdW) trilayers via intercalating a two-dimensional (2D) monolayer (ML = BN, MoSe$_2$, WS$_2$, or WSe$_2$) between MoS$_2$ bilayer to form various MoS$_2$/ML/MoS$_2$ sandwich trilayers. We find that the BN monolayer is the most effective sheet to decouple the interlayer vdW coupling of the MoS$_2$ bilayer, and the resulting sandwich trilayer can recover the electronic structures of the MoS$_2$ monolayer, particularly the direct-gap character. Further study of the MoS$_2$/BN superlattices confirms the effectiveness of the BN monolayer for the decoupling of the MoS$_2$-MoS$_2$ interaction. In addition, the intercalation of transition-metal dichalcogenide (TMDC) MoSe$_2$ or WSe$_2$ sheet renders the sandwich trilayer undergoing an indirect-gap to direct-gap transition due to the newly formed heterogeneous S/Se interfaces. In contrast, the MoS$_2$/WS$_2$/MoS$_2$ sandwich trilayer still retains the indirect-gap character of the MoS$_2$ bilayer due to the lack of the heterogeneous S/Se interfaces. Moreover, the 3D superlattice of the MoS$_2$/TMDC heterostructures also exhibits similar electronic band characters as the MoS$_2$/TMDC/MoS$_2$ trilayer heterostructures, albeit slight decrease of the bandgap than the trilayers. Compared to the bulk MoS$_2$, the 3D MoS$_2$/TMDC superlattice can give rise to new and distinctive properties. Our study offers not only new insights into electronic properties of the vdW multilayer heterostructures but also guidance in designing new heterostructures to modify electronic structures of 2D TMDC crystals.

**Introduction**

Many experiments have demonstrated that two-dimensional (2D) transition-metal dichalcogenides (TMDCs) such as 2D $MoS_2$ and $WS_2$ crystals possess novel electronic,[1-4] optical,[5-8] catalytic,[9,10] and mechanical properties.[11-14] For example, electronic properties of 2D $MoS_2$ crystals can be sensitive to the number of layers, that is, the $MoS_2$ monolayer exhibits a direct bandgap with a value of ~1.8 eV while a bilayer $MoS_2$ exhibits an indirect bandgap with a value of ~1.5 eV.[15] As a result, significant enhancement of photoluminescence has been observed when $MoS_2$ is thinned to a single layer.[8,14] Previous study has also shown that when the $MoS_2$ bilayer is pulled apart into two separated monolayers, the direct transition (K-K) is insensitive to the separation while the indirect transition (Γ-K) increases dramatically.[16] It seems that the distance between the two monolayers or the interlayer vdW interaction can notably affect the electronic structures of two-dimensional (2D) layered TMDCs. Thus, one may ask two closely related questions: "Can the $MoS_2$ bilayer be effectively decoupled via intercalation of a 2D sheet without being pulled far too apart?" or "To what extent, can the intercalation of a 2D sheet affect electronic properties of the $MoS_2$ bilayer?" The intercalation of a 2D sheet into the $MoS_2$ bilayer gives rise to a hybrid trilayer, coined as the vdW heterostructures by Geim and Grigorieva.[17] Recently, successful fabrication of multilayer vdW heterostructures by stacking one layer on top of another in a precisely controlled sequence has been demonstrated experimentally.[18-20] The artificial vdW heterostructures can exhibit new and unusual properties that differ from their own constituent layers. For example, the vertical field-effect transistor and memory cell made of TMDC/graphene heterostructures[19,21-23] as well as layered hybrids of $MoS_2$ and $WS_2$ have been realized in the laboratory.[24] Previous theoretical studies suggest that the direct-gap character of the MoS$_2$ monolayer can be retained in certain MoS$_2$ heterobilayer structures whose electronic properties can be further tuned by an in-plane strain or a vertical electric field.[25-29] In addition, the insulating BN monolayer is a good substrate for protecting high quality graphene electronics.[30] A type-I band alignment for BN monolayer and MoS$_2$ monolayer is also reported.[25] 3D heterostructures such as superlattices are predicted to possess new properties that differ from the corresponding bulk structures, thereby opening a new way of materials design.[31]

The focus of this study is to investigate effects of intercalation of either an insulating BN monolayer or a semiconducting TMDC monolayer (MoSe$_2$, WS$_2$, or WSe$_2$) into MoS$_2$ bilayer on the electronic properties of the vdW trilayer heterostructure and the corresponding vdW superlattice. Our computational results suggest that the BN monolayer is an ideal sheet to decouple the MoS$_2$ bilayer while MoSe$_2$ or WSe$_2$ sheet can turn the indirect-gap of MoS$_2$ bilayer into a direct-gap trilayer.

**Computational Methods:**

All calculations are performed within the framework of spin-polarized plane-wave density functional theory (PW-DFT), implemented in the Vienna ab initio simulation package (VASP 5.3).[32, 33] The Perdew–Burke–Ernzerhof (PBE) functional and projector augmented wave (PAW) potentials are used.[34-36] Effect of vdW interaction is accounted for by using the dispersion corrected DFT ( optB88-vdW functional).[37, 38] The vacuum length between two adjacent images in the supercell is longer than 15 Å. An energy cutoff of 500 eV is adopted for the plane-wave expansion of the electronic wave function. Geometric structures are relaxed until the force on each atom is less than 0.01 eV/Å and the convergence criteria for energy is $1 \times 10^{-5}$ eV.

Note that the optimized MoS$_2$ monolayer exhibits a cell parameter of 3.18 Å, while the cell parameter of $h$-BN monolayer is 2.52 Å, in good agreement with previous results.[39, 40] As such, the 5×5 BN supercell almost perfectly matches the 4×4 MoS$_2$ supercell with the lattice mismatch less than 1%. For the MoS$_2$/BN/MoS$_2$ trilayer, the supercell is fixed while the atomic coordinates are relaxed only. For other sandwich systems containing TMDC $MoSe_2$, $WS_2$, or $WSe_2$, a 1×1 cell is used due to their lattice parameters are close to that of $MoS_2$. In these cases, both the cell length and atomic coordinates are relaxed to obtain the lattice parameters at the lowest total energy. Bader's atom in molecule (AIM) method based on charge density topological analysis is used for computing charge population.[41] Once the optimized structures are obtained, a hybrid functional in the Heyd-Scuseria-Ernzerhof (HSE06) form is used to give more accurate bandgaps.[42] The spin-orbit (SO) interaction is included in all the band-structure calculations except the HSE06 band-structure calculations for $MoS_2$/BN/$MoS_2$ trilayer with A1B1A1 stacking (see below).[43]

**Results and Discussion**

**1. Intercalation of $MoS_2$ Bilayer with BN Monolayer**

Frist, DFT/PBE calculations show that monolayer $MoS_2$ is a semiconductor with a direct bandgap of 1.60 eV (see Table 1), in agreement with previous calculation.[29] $MoS_2$ bilayer with the most stable C7 stacking, however, is a semiconductor with an indirect bandgap of 1.31 eV. HSE06 calculation enlarges the bandgap of monolayer and bilayer $MoS_2$ to 2.06 and 1.81 eV, respectively. For the BN monolayer, PBE calculation shows it is a semiconductor with a wide bandgap of 4.66 eV.

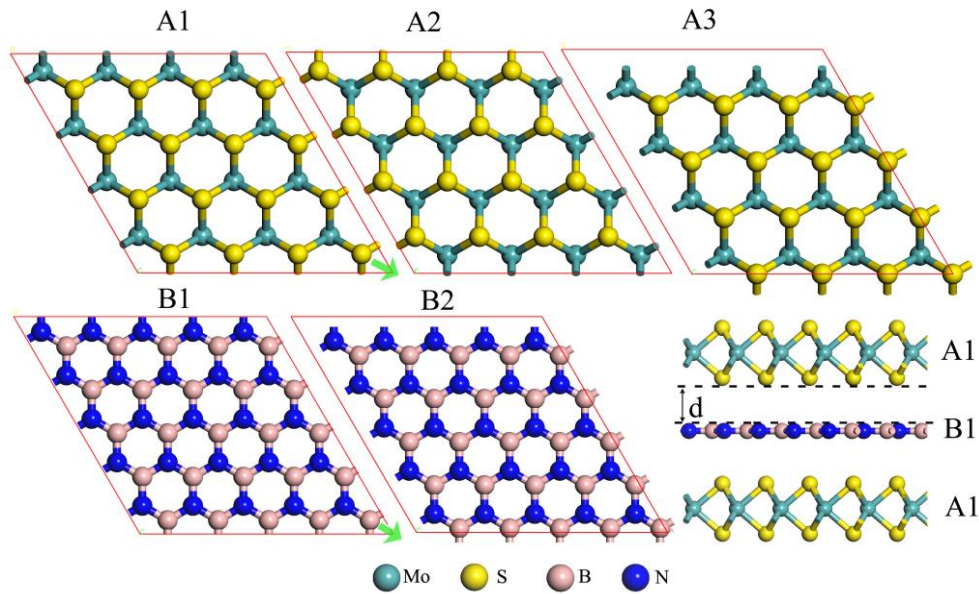

**Figure 1.** Top view of a MoS$_2$ monolayer in three different supercells (marked by the red parallelogram) and a BN monolayer in two different supercells. A3 (B2) can be viewed as a displacement of A1 (B1) in the green-arrow direction shown in A1 (B1). Superimposing the ABA supercells allows us to build different MoS$_2$/BN/MoS$_2$ trilayers. An example of A1B1A1 trilayer is shown in the lower right panel.

**Table 1**. The distance $d$ (in Å) between two nearest-neighbor monolayers as shown in Figure 1. The binding energy E$_{BE}$ (in eV) per formula unit. The computed bandgaps E$_g$ (in eV) of MoS$_2$ monolayer (ML-MoS$_2$), bilayer (BL-MoS$_2$), trilayer (MoS$_2$/BN/MoS$_2$) heterostructures and related superlattice (SL) with different stacking orders. The SO effect is included in HSE06 calculation of the bandgap except the largest trilayer system A1B1A1.

|  | ML-MoS$_2$ | BL-MoS2 | A1B1A1 | A1B2A1 | A1B1A2 | A1B1A3 | SL-A1B1 | SL-A1B2 |
|---|---|---|---|---|---|---|---|---|
| $d$ | / | 3.09 | 3.36 | 3.36 | 3.36 | 3.36 | 3.35 | 3.35 |
| E$_{BE}$ | / | 0.22 | 0.36 | 0.36 | 0.37 | 0.36 | 0.38 | 0.38 |
| Eg(PBE) | 1.60 | 1.31-indirect | 1.58 | 1.58 | 1.58 | 1.58 | 1.69 | 1.69 |
| Eg(HSE06) | 2.06 | 1.81-indirect | 2.11 | / | / | / | / | / |

Next, various MoS$_2$/BN/MoS$_2$ trilayer systems are built for which the lateral locations of the MoS$_2$ monolayer and BN monolayer in different supercells are shown in Figure 1. Specifically, we consider four different stacking orders: A1B1A1, A1B2A1, A1B1A2, and A1B1A3. PBE optimizations show the total-energy differences among these configurations is typically less than 0.01 eV per formula cell, and the different stacking orders have little effect on the electronic structures. The binding energy of a trilayer, which measures the interlayer vdW interaction per supercell, is defined as: E$_{BE}$ = 2E$_{MoS2}$ + E$_{BN}$ − E$_{MoS2/BN/MoS2}$, where E$_{MoS2}$ is the total energy of a MoS$_2$ monolayer, E$_{BN}$ is the total energy of a BN monolayer, and E$_{MoS2/BN/MoS2}$ is the total energy of a MoS$_2$/BN/MoS$_2$ trilayer. As listed in Table 1, the computed binding energy of MoS$_2$/BN/MoS$_2$ heterostructure with A1B1A1, A1B2A1, A1B1A2, A1B1A3 stacking orders are 0.36 eV, 0.36 eV, 0.37 eV, and 0.36 eV, respectively, reflecting the weak vdW interaction between the MoS$_2$ layer and BN layer. Taking the A1B1A1 stacking as an example, its electronic structure is shown in Figure 2c. Clearly, the trilayer retains the direct-gap character of the $MoS_2$ monolayer. The computed bandgap is 1.58 eV, and both the conduction band minimum (CBM) and valence band maximum (VBM) are located at the K points, both contributed by the $MoS_2$ layers. Like the PBE calculation, the HSE06 calculation also suggests direct-gap character but the bandgap increases to 2.11 eV (ESI Figure S1). Overall, the intercalated BN layer has little effect on the band edge of $MoS_2$ layers. To further confirm this conclusion, we remove the BN layer but leave the two $MoS_2$ layers fixed at the original locations of the trilayer. As shown in Figure 2b, again, the computed band structure shows direct-gap character with the bandgap being 1.59 eV, very close to that of $MoS_2/BN/MoS_2$ trilayer.

We also compute the effective mass at the K point corresponding to the $MoS_2$ monolayer and A1B1A1 trilayer, respectively. The directional dependence of the effective mass at the K point is small. For $MoS_2$ monolayer, $m^*_e=0.44\ m_0$ for the electron at CBM, and $m^*_h=0.55\ m_0$ for the hole at VBM, are in agreement with the previous studies.[4,44] For A1B1A1, $m^*_e=0.44\ m_0$ and $m^*_h=0.59\ m_0$, similar to that of the monolayer, which indicates the carrier mobility of the monolayer is also retained by the trilayer.

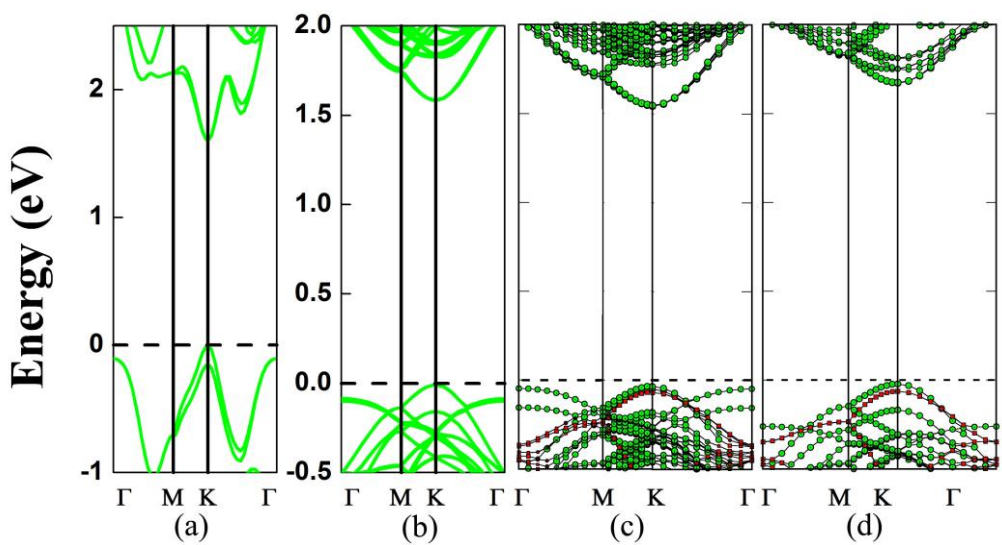

**Figure 2.** Computed electronic band structures (PBE) of (a) monolayer $MoS_2$; (b) $MoS_2$/vacuum layer/$MoS_2$ by removing the BN layer from the $MoS_2/BN/MoS_2$ trilayer

counterpart (in (c)) but with the fixed vertical location of the two MoS$_2$ layers; (c) MoS$_2$/BN/MoS$_2$ with the A1B1A1 stacking; and (d) a 3D superlattice of MoS$_2$/BN with the AB stacking. The green lines represent MoS$_2$ layers while the red lines represent BN monolayer.

To confirm that the BN monolayer is an ideal sheet to decouple the interlayer coupling of MoS$_2$ bilayer, we also compute electronic structures of the 3D superlattice of hybrid BN/MoS$_2$ layers. Superlattice with two different stacking orders (A1B1 and A1B2) of MoS$_2$ and BN layers is considered and our calculations show the two stacking orders give nearly the same results. For both stacking orders, the optimized cell parameters are $a = b = 12.62$ Å and c = 9.86 Å. As shown in Figure 2d, the superlattice exhibits a direct gap of 1.69 eV, and both the VBM and CBM are located at the K point and both are contributed by MoS$_2$ layers as in the case of the MoS$_2$/BN/MoS$_2$ trilayer system. The slightly enhanced bandgap compared to the trilayer system is largely due to slight reduction of the cell parameters *a* and *b*. In summary, results of both vdW trilayer and superlattice show that the alternatively stacked BN and MoS$_2$ monolayers can retain the direct-gap character of the MoS$_2$ monolayer. In other words, BN monolayer is an effective divider to decouple the interlayer coupling for MoS$_2$.

**2. Intercalation MoS$_2$ Bilayer by MoSe$_2$, WS$_2$ or WSe$_2$ Monolayer**

Previous experimental and theoretical studies demonstrate that MoS$_2$ bilayer is a semiconductor with an indirect bandgap.[27, 29, 45] Recent theoretical studies of TMDC heterobilayers also show that the interlayer interaction due to hetero interface (e.g., S/Se) can notably affect the electronic properties. Thus, it is interesting to study the extent to which the intercalation of a heterogeneous TMDC monolayer between two MoS$_2$ bilayers affects the electronic properties.

Previous theoretical studies show that the C7 and T stacking patterns give the lowest energy for many heterobilayers,[27, 29] and the electronic structure is more or less the same with different stacking orders. Here, we adopt two different stacking orders for the trilayers (see Figure 3), namely, the ABA and ACA. For the ABA trilayer, the interface AB is in C7 stacking, while for the ACA trilayer, the interface AC is in T stacking. Again, we find that the two stacking orders give rise to nearly the same electronic properties.

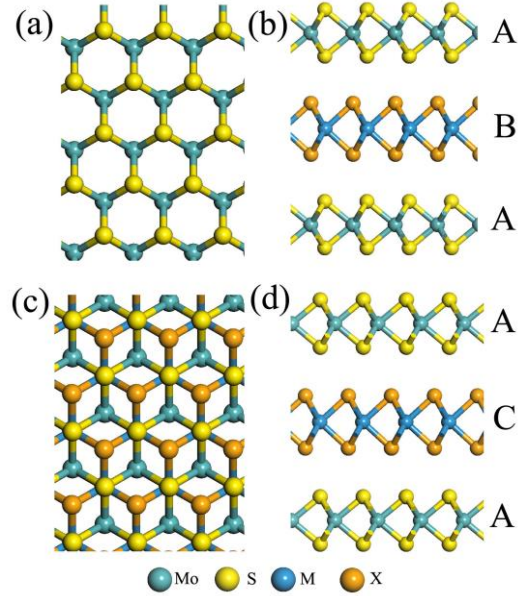

**Figure 3**. Top and side views of $MoS_2/MX_2/MoS_2$ (M=Mo, W; X=S, Se) trilayers with (a) and (b) ABA stacking with C7 interface, (c) and (d) ACA stacking with T interface, respectively.

**Table 2**. Computed PBE $E_g$(PBE) and HSE06 $E_g$(HSE) bandgaps of $MoS_2/ML/MoS_2$ trilayers (ML = $MoSe_2$, $WS_2$ or $WSe_2$) in ABA and ACA stacking. The PBE ($E_g$(PBE)_SL) and HSE06 ($E_g$(HSE)_SL) bandgaps of $MoS_2/ML$ superlattice (ML = $MoSe_2$, $WS_2$, $WSe_2$) in AB and AC stacking. The unit is in eV.

|  | ABA | ACA | ABA | ACA | ABA | ACA |
|---|---|---|---|---|---|---|
|  | (B=$MoSe_2$) | (C=$MoSe_2$) | (B=$WS_2$) | (C=$WS_2$) | (B=$WSe_2$) | (c=$WSe_2$) |
| Eg(PBE) | 0.70 | 0.75 | 1.05 (Γ-k) | 1.08 (Γ-k) | 0.39 | 0.42 |
| Eg(HSE) | 0.97 | 1.02 | 1.47 (Γ-k) | 1.49 (Γ-k) | 0.61 | 0.64 |
| Eg(PBE)_SL | 0.62 (Γ-k) | 0.59 (Γ-k) | 0.95 (Γ-k) | 0.93 (Γ-k) | 0.30 | 0.35 |
| Eg(HSE)_SL | 0.88 | 0.92 | 1.35 (Γ-k) | 1.33 (Γ-k) | 0.50 | 0.54 |

The polarization within the S/Se interfaces is responsible to the direct-gap character for heterobilayers in previous studies.[26, 27] The MoS$_2$/MoSe$_2$/MoS$_2$ trilayer entails two S/Se interfaces. In view of the MoS$_2$ bilayer possessing an indirect bandgap, the intercalation of MoSe$_2$ monolayer induces an indirect to direct transition. As shown in Figure 4a, the MoS$_2$/MoSe$_2$/MoS$_2$ trilayer exhibits a direct bandgap of 0.69 eV. The VBM is located at the K point and is mainly contributed by the MoSe$_2$ layer; while the CBM is also located at the K point and is mainly contributed by MoS$_2$ layers. It is desirable that CBM and VBM are contributed from two different TMDC monolayers, particularly for the electron-hole separation. Electronic Supplemental Information (ESI Figure S2) shows a more accurate HSE06 computation of band structures of the trilayers. The computed bandgap is 0.97 eV, wider than that from PBE computation. However, the electronic structures and the VBM and CBM are similar one another for PBE computation.

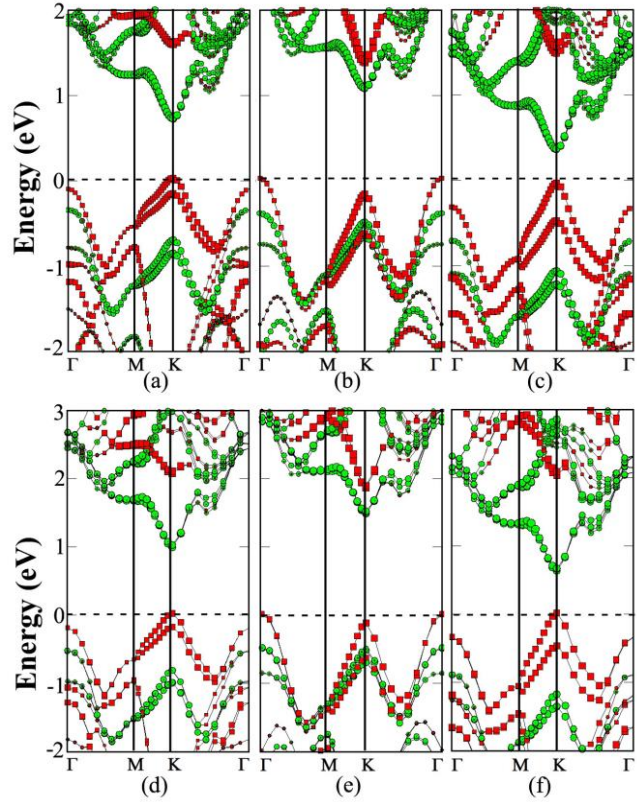

**Figure 4.** Computed band structures (PBE) of (a) MoS$_2$/MoSe$_2$/MoS$_2$, (b) MoS$_2$/WS$_2$/MoS$_2$, and (c) MoS$_2$/WSe$_2$/MoS$_2$ trilayer with ABA stacking, respectively. Computed band structures (PBE) of (d) MoS$_2$/MoSe$_2$ superlattice, (e) MoS$_2$/WS$_2$ superlattice, and (f) MoS$_2$/WSe$_2$ superlattice with the AB stacking, respectively. The green lines mark contribution from MoS$_2$ layers while the red lines mark contribution from MoSe$_2$, WS$_2$ or WSe$_2$ layer.

Because of the lack of the S/Se interfaces, as shown in Figure 4b, the MoS$_2$/WS$_2$/MoS$_2$ trilayer still exhibits an indirect gap of 1.05 eV. The VBM is located at the Γ point and is mainly contributed by the WS$_2$ monolayer, while the CBM is located at the K point and is mainly contributed by the two MoS$_2$ layers. Computed band structures based on the HSE06 functional is shown in ESI Figure S2. Again, the trilayer still exhibits the indirect-gap character but the bandgap increases to 1.47 eV.

The MoS$_2$/WSe$_2$/MoS$_2$ trilayer still exhibits a direct gap of 0.39 eV due to the presence of the two Se/S interfaces (see Figure 4c). The VBM at the K point is mainly contributed by the WSe$_2$ layer, while the CBM at the K point is mainly contributed by MoS$_2$ layers. Again, as shown in ESI Figure S2, HSE06 calculation confirms the main character of electronic properties. To further analyze the effect of the polarization within the Se/S interfaces, charge transfer between neighboring layers is computed (see ESI Table S1). For MoS$_2$/MoSe$_2$/MoS$_2$ and MoS$_2$/WSe$_2$/MoS$_2$ trilayers, a 0.02 $e$ per unit cell is transferred from MoSe$_2$ layer to MoS$_2$ layer. In contrast, for MoS$_2$/WS$_2$/MoS$_2$ trilayer with S/S interfaces, the charge transfer between two neighboring layers is nearly zero. This result further demonstrates that the interfacial polarization has an important effect on the electronic properties of the trilayer heterostructures.

Lastly, we consider 3D superlattice made of hybrid MoS$_2$ monolayers and another monolayers. As shown in Figure 4, for each superlattice, two stacking orders including AB with C7 interface and AC with T interface are investigated. The binding energies and cell parameters for the AB and AC stacking are close to one another in all the configurations (see ESI Table S1). For the MoS$_2$/MoSe$_2$ superlattice, PBE calculations suggest that its bandgap is still indirect, with a value of 0.62 and 0.59 eV respectively, for the AB and AC stacking. The bandgap is about 0.1 eV less than that of the corresponding trilayer. The CBM is still located at the K point and contributed mainly by the MoS$_2$ layers (Figure 4d), while the VBM energy at the Γ and K point differs only by 10 meV, and is mainly contributed by the MoSe$_2$ layers. On the other hand, the HSE06 calculations suggest that the MoS$_2$/MoSe$_2$ superlattice is a direct-gap semiconductor with a value of 0.88 and 0.92 eV, respectively, for the AB and AC stacking. Here, the VBM energy in the K point is 77 meV lower than the Γ point (ESI Figure S2). For MoS$_2$/WS$_2$ superlattice, both PBE and HSE06 calculations suggest that it is an indirect-gap semiconductor (Figure 4e) with a value 0.95 and 1.35 eV, respectively, for the AB stacking. Finally, both PBE and HSE06 calculations suggest that the MoS$_2$/WSe$_2$ superlattice is a direct-gap semiconductor with a value of 0.3 and 0.5 eV, respectively, for the AB and AC stacking. Both bandgaps are smaller than those of the corresponding trilayers. Again, the bandgap reduction is mainly due to slightly enlarged cell parameter (see ESI Table S1).

## 3. Conclusions

In conclusion, our first-principles calculations show that the BN monolayer is a highly effective single sheet to decouple the interlayer vdW interaction of the MoS$_2$ bilayer. The resulting vdW trilayer heterostructure can recover the electronic structures of a single MoS$_2$ monolayer, particularly its direct-gap character. Further study of the 3D MoS$_2$/BN superlattices confirms the effectiveness of the BN monolayer for decoupling the interlayer interaction. Expectedly, this conclusion has implications to MoS$_2$ based heterostructures as well as to other TMDC-based vdW heterostructures. We have also investigated intercalation of a TMDC MoSe$_2$ or WSe$_2$ sheet between two MoS$_2$ sheets and found that the resulting vdW trilayer undergoes an indirect-gap to direct-gap transition due to the newly formed heterogeneous S/Se interfaces. In contrast, the MoS$_2$/WS$_2$/MoS$_2$ vdW trilayer still retains the indirect-gap character of the MoS$_2$ bilayer due to the lack of the heterogeneous S/Se interfaces. Again, the 3D superlattice of the MoS$_2$/TMDC heterostructures also exhibits similar electronic band characters as the MoS$_2$/TMDC/MoS$_2$ trilayer, albeit slight decrease of the bandgap than that of the trilayer counterparts. In view of recent successful fabrication of vdW heterostructures by stacking a graphene sheet on top of MoS$_2$ sheets or *vice versa*,[19] the vdW trilayers and superlattices investigated in this study together with their novel properties may be tested in the laboratory in near future.

**Acknowledgements**

XCZ is grateful to valuable discussions with Professors Ali Adibi, Eric Vogel, Joshua Robinson, and Ali Eftekhar. The USTC group is supported by the National Basic Research Programs of China (Nos. 2011CB921400, 2012CB 922001), NSFC (Grant Nos. 21121003, 11004180, 51172223), One Hundred Person Project of CAS, Strategic Priority Research Program of CAS (XDB01020300). UNL group is supported by ARL (Grant No. W911NF1020099), NSF (Grant No. DMR-0820521), UNL Nebraska Center for Energy Sciences Research, University of Nebraska Holland Computing Center, and a grant from USTC for (1000plan) Qianren-B summer research.

TOC graphic

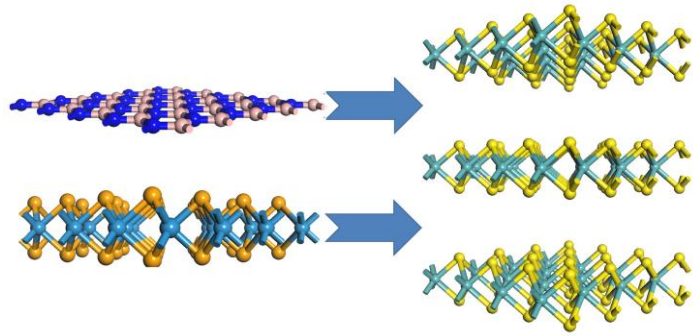